\documentclass[10pt,letterpaper]{article}
\usepackage{opex3}
\usepackage{cite}

\begin{document}


\vskip4pc


\title{The biaxial nonlinear crystal \textrm{BiB$_3$O$_6$} as a polarization entangled photon source using non-collinear type-II parametric down-conversion}

\author{A. Halevy,$^1$ E. Megidish,$^1$, L. Dovrat,$^1$ H. S. Eisenberg,$^1$ P. Becker,$^2$ and L. Bohat\'{y}$^2$}

\address{$^1$Racah Institute of Physics, Hebrew University of Jerusalem, \\ Jerusalem 91904, Israel\\ $^2$Institute of Crystallography, University of Cologne, \\ 50939 Cologne, Germany}
\email{hagaie@huji.ac.il} 


\begin{abstract}
We describe the full characterization of the biaxial nonlinear
crystal \textrm{BiB$_3$O$_6$} (BiBO) as a polarization entangled
photon source using non-collinear type-II parametric
down-conversion. We consider the relevant parameters for crystal
design, such as cutting angles, polarization of the photons,
effective nonlinearity, spatial and temporal walk-offs, crystal
thickness, and the effect of the pump laser bandwidth. Experimental
results showing entanglement generation with high rates and a
comparison to the well investigated $\beta$-\textrm{BaB$_2$O$_4$}
(BBO) crystal are presented as well. Changing the down-conversion
crystal of a polarization entangled photon source from BBO to BiBO
enhances the generation rate as if the pump power was increased by
2.5 times. Such an improvement is currently required for the
generation of multiphoton entangled states.
\end{abstract}



\section{Introduction}

For more than two decades, parametric down-conversion (PDC) is a
central tool for the generation of entangled photons. This is a
second order nonlinear process, where a photon from a pump beam
splits into two photons, known as \emph{signal} and \textit{idler},
while conserving energy and momentum. The down-converted photons
exhibit strong correlations in various degrees of freedom, such as
wavelength, time of emission, polarization, momentum, and position
\cite{Ghosh87}. The down-conversion pair generation rate depends
linearly on the pump beam power, and quadratically on the crystal
thickness and on its nonlinear coefficients \cite{Boyd03}. Since the
first demonstration of an efficient PDC polarization entangled
photon source \cite{Kwiat95}, the most commonly used nonlinear
birefringent crystal for this purpose is the uniaxial crystal
$\beta$-\textrm{BaB$_2$O$_4$} (BBO). The reasons for this are its
relatively high nonlinear coefficients, high transparency, and the
possibility for phase-matching over a broad spectral window
\cite{Becker98, Eckardt90}.

In the last decade, many quantum optics experiments have used two
consequent PDC events \cite{Bouwmeester97, Pan98, Pan01}. Others
have used second order events of PDC \cite{Bouwmeester99,
Lamas-Linares01, Halevy11}. These events occur when two
indistinguishable pump photons split into four, during the same
coherence time (or pulse duration for pulsed pump sources). Both
approaches require high efficiency of the PDC process as success
probability is quadratic with the single pair generation
probability. Later, the third order PDC event as well as three
consequent first order events have been used to create entangled
states of six photons \cite{Lu07, Radmark09, Wagenknecht10, Barz10}.
Recently, four consequent first order PDC events were used to
demonstrate an eight photon entangled state \cite{Yao11}.

One possibility to enhance the PDC generation probability is to use
thicker nonlinear crystals. The crystal length is limited by the
non-collinearity of the process, that spatially separates the pump
beam and the down-converted photons. In addition, there is the
spatial walk-off effect between the two polarizations that degrades
the entanglement quality (see Sec. \ref{subsec:SWO}). Thus, higher
pump intensity is required. Usually, the pump beam is generated by
frequency doubling the radiation of a Ti:Sapphire laser in another
nonlinear crystal \cite{Bouwmeester97, Pan98, Pan01, Bouwmeester99,
Lamas-Linares01, Halevy11, Lu07, Radmark09, Wagenknecht10, Barz10,
Yao11}. Reported typical intensities are above 1\,W, but as the
doubling crystal is damaged by the high power, it has to be
translated continuously in order to maintain stable operation
\cite{Lu07}. Additionally, the pump beam intensity can be enhanced
inside a synchronized external cavity. Such a setup has been shown
to pump a BBO crystal with about 7\,W \cite{Krischek10}.

In this work, we suggest and demonstrate the use of a novel crystal
with higher nonlinear coefficients than BBO for the generation of
polarization entangled photons. It is the monoclinic biaxial
\textrm{BiB$_3$O$_6$} (BiBO) crystal that has been introduced
\cite{Bohaty99} and characterized \cite{Bohaty00} as a nonlinear
optical crystal about a decade ago. Since then, it was used in
numerous frequency conversion experiments (for example, see Ref.
\cite{Petrov09}, and Refs. within). BiBO was also used with type-I
PDC for generating photon pairs with a pulsed laser source
\cite{Higgins07} and for generating polarization entangled photons
with a continuous pump source \cite{Rangarajan09}. It has a very
broad transparency window and its nonlinear coefficients are
considerably higher than those of BBO \cite{Bohaty99}. Nevertheless,
the biaxiallity introduces many differences and difficulties,
compared to BBO.

This paper is organized as follows: in Sec. \ref{sec:PDCParamaters}
we present the various considerations in choosing the crystal
parameters. These parameters are affected by the phase-matching
angles, the polarization direction of the pump beam and the
down-converted photons, the pump beam bandwidth, the spectral and
angular properties of the down-converted photons, spatial and
temporal walk-off effects, and the effective second order nonlinear
coefficient $d_{eff}$ dependence on the pump beam direction. Section
\ref{sec:ExperimentalResults} describes the experimental validation
of our theoretical results by demonstrating and quantifying the
entanglement produced by using two known configurations.

\section{Investigation of PDC parameters in BiBO}
\label{sec:PDCParamaters}
\subsection{Crystal design}

In order to lower reflections and to simplify the required
calculations and alignment, it is desirable to cut the crystal
facets perpendicular to the designed direction of the wave vector
$\textbf{k}_f$ of the fundamental (pump) wave. The phase-matching
calculation for the $\textbf{k}_f$ direction scans a quadrant of
space, according to the monoclinic symmetry of BiBO. In order to
choose the optimal phase-matching direction, the effective second
order nonlinear coefficient $d_{eff}$ is calculated for each
$\textbf{k}_f$ direction. As an approximation for $d_{eff}$, we use
the effective nonlinear coefficient of collinear second-harmonic
generation, $d_{eff}^{SHG}$, calculated for any direction, even
though the phase-matching condition is not fulfilled. The optimum
direction of $\textbf{k}_f$ within the range of the highest values
of $d_{eff}$ should allow the two cones to intersect at $90^\circ$,
which is optimal for the photon collection efficiencies. For the
selected direction of $\textbf{k}_f$ as well as for the
down-converted photons at the intersection points of the emission
cones, the polarization orientation is calculated. Finally, the
temporal and the spatial walk-offs are calculated for the chosen
crystal parameters.

\subsection{Phase-matching calculation}

\begin{figure}[tb]
\centering\includegraphics[width=5in]{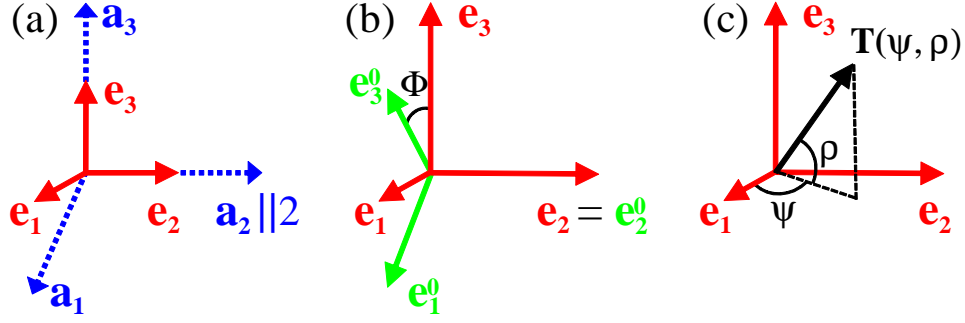}
\caption{\label{ReferenceFrames} \textbf{(a)} Relative orientation
of the crystallographic axes $\{a_i\}$ and the crystal physical axes
$\{e_i\}$. \textbf{(b)} Orientation of the crystal physical axes
$\{e_i\}$ and the optical indicatrix main axes $\{e_i^0\}$. $\Phi$
is the angle of orientational dispersion of the principal axes.
\textbf{(c)} The propagation direction of the pump beam \textbf{T}
inside the crystal is defined using the two angles $\psi$ and $\rho$
in the wavelength independent $\{e_i\}$ system.}
\end{figure}

\begin{figure}[tb]
\centering\includegraphics[width=5in]{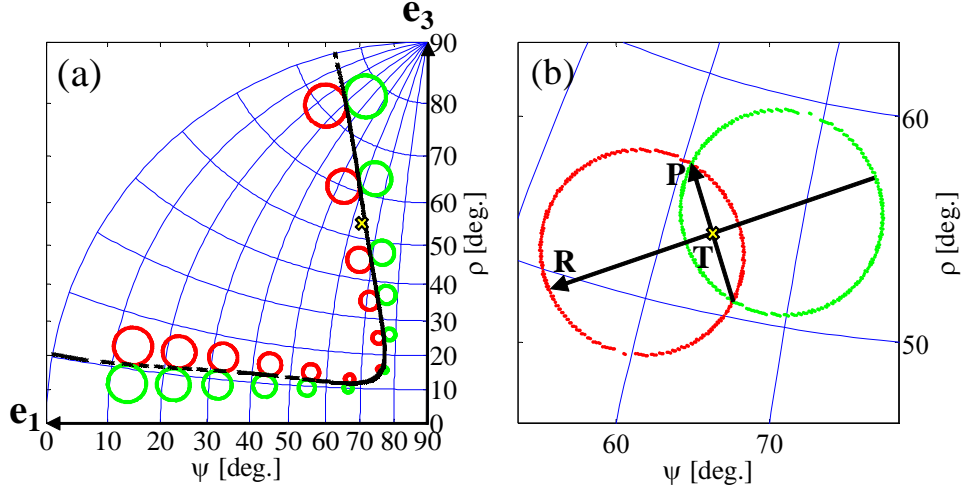}
\caption{\label{StreographicProjection} \textbf{(a)} Stereographic
projection of collinear type-II phase-matching angles for
$\lambda_f$=390\,nm (black line), with several non-collinear
down-converted circles for different propagation directions of the
fundamental wave. The chosen working point for this work is marked
with X. \textbf{(b)} Stereographic projection of PDC in BiBO for
$\psi(\textbf{T})=63.5^\circ$ and $\rho(\textbf{T})=53.5^\circ$.
The vectors \textbf{P} and \textbf{R} are indicated, as defined in
the text.}
\end{figure}

In order to find the spatial distribution of the cones of
down-converted photons, we calculated numerically the
non-collinear type-II PDC process in BiBO. We are interested in
the degenerated case in which the down-converted photons share the
same wavelength. The most basic reference system that we use is
the crystal physical Cartesian system $\{e_i\}$. It is linked to
the crystallographic system $\{a_i\}$ (see Ref. \cite{Frohlich84})
by $e_3
\parallel a_3, e_2\parallel a_2 \parallel$ 2-fold axis, $e_1 =
e_2 \times e_3$, see Fig. \ref{ReferenceFrames}(a). The point group
symmetry 2 of the monoclinic BiBO crystal structure allows the
occurrence of enantiomorphic (i.e., "left-handed" and
"right-handed") species. All our samples for optical investigations
were prepared using crystals that were grown as descendants from the
same parent crystal and therefore posses the same handedness. For
our crystals, the positive direction of $a_2$ (and $e_2$)
corresponds to a positive sign of the pyroelectric coefficient
$p^\sigma_2$ (at constant stress) and to a negative sign of the
longitudinal piezoelectric coefficient $d_{222}$ \cite{IEEE87,
Haussuhl06}.

In BiBO, the principal axes $\{e_i^0\}$ of the optical indicatrix
coincide with the $\{e_i\}$ system only for $e_2^0=e_2$ while
$e_1^0$ and $e_3^0$ change their orientation with wavelength. This
orientational dispersion is illustrated by the angle
$\Phi=\angle(e_3, e_3^0)$ in Fig. \ref{ReferenceFrames}(b). For the
fundamental and the down-converted wavelengths used in this work
($\lambda_f=390$\,nm and $\lambda_{dc}=780$\,nm), $\Phi$ equals
$43.8^\circ$ and $46.9^\circ$, respectively \cite{Bohaty00}.

Our calculations of the collinear and non-collinear PDC
phase-matching cases \cite{Hobden67}, basically follows the
calculation strategy described by Ref. \cite{Boeuf00}. For a chosen
direction \textbf{T} of the fundamental wave vector $\textbf{k}_f$,
we define the propagation direction in spherical coordinates ($\psi,
\rho$) with respect to $\{e_i\}$ (see Fig.
\ref{ReferenceFrames}(c)). The phase-matching conditions are
satisfied when
\begin{equation}\label{PhaseMatching}
\Delta\textbf{k}=\textbf{k}_{signal}+\textbf{k}_{idler}-\textbf{k}_f=0,
\end{equation}
where $\Delta\textbf{k}$ is the phase-mismatch vector, and
$\textbf{k}_{signal}$ and $\textbf{k}_{idler}$ are the wave vectors
of the down-converted waves. First, we find the collinear
phase-matching angles, as in this case Eq. \ref{PhaseMatching}
becomes scalar and simple to solve. Then, we use a search algorithm
around the collinear direction to find the non-collinear directions
that correspond to the minimal values of $\Delta\textbf{k}$. We have
chosen a numerical threshold value of
$|\frac{\Delta\textbf{k}}{\textbf{k}_f}|<5\times10^{-5}$. Photons
are emitted into two cones with different, and not necessary
perpendicular, polarizations. The stereographic projections of
several down-converted emission cones onto the ($e_1, e_3$) plane
are presented in Fig. \ref{StreographicProjection}(a). This
projection preserves angles and projects circles in three dimensions
as circles on the plane \cite{Hobden67}. Each two tangent circles
represent a non-collinear solution, where the direction of the
fundamental wave $\textbf{k}_f$ is their collinear intersection
point. The down-converted photons experience refraction when they
emerge from the crystal to air, which depends on their propagation
direction and their polarization. The calculation results given in
this work are of the photon's properties outside the crystal. For
our wavelength parameters, the phase-matching calculations resulted
in a suitable direction \textbf{T} with spherical coordinates $\psi=
63.5^\circ$ and $\rho=53.5^\circ$. In this case, the two
down-converted cones intersect at an angle of $90^\circ$ and the
intersection points are separated by $6.9\pm0.2^\circ$. In order to
simplify the crystal alignment process, it is convenient to define a
sample reference system according to the PDC emission results. The
direction of the wave vector $\textbf{k}_f$ of the fundamental wave
is parallel to \textbf{T}. \textbf{T} is also normal to the input
facet of the sample. We define \textbf{P} to be the vector
connecting the two cones intersection points (see Fig.
\ref{StreographicProjection}(b)) and \textbf{R} the vector that
connects the most distant points on each circle. Consequently,
\textbf{T}, \textbf{P}, and \textbf{R} form an orthogonal set. The
BiBO samples used in our PDC experiments have spherical coordinates
$(\psi,\rho)$ of $\textbf{T}=(63.5^\circ, 53.5^\circ)$,
$\textbf{P}=(-80.6\pm0.1^\circ, +30.9\pm0.1^\circ)$, and
$\textbf{R}=(-1.4\pm0.1^\circ, -17.4\pm0.1^\circ)$. The errors
result from the finite grid resolution of the calculation for the
intersection points.

\subsection{The photons' polarization}
\label{subsec:Polarization}

\begin{figure}[tb]
\centering\includegraphics[width=5in]{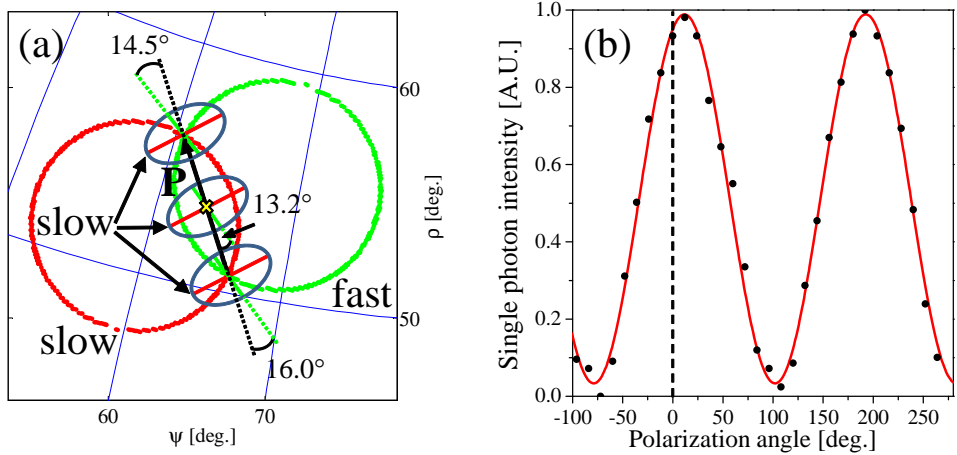}
\caption{\label{BiBOPolarizationModes} \textbf{(a)} The relevant
photon polarization directions for the designed crystal. The
elliptical cross-sections of the wavelength dependant indicatrix are
marked for the fundamental beam and the two cones intersection
directions. Note that the ellipticity of the cross-sections is
exaggerated for clarity reasons. The long and short semi-axes of the
cross-sections indicate the polarization directions of the slow and
fast waves, respectively. \textbf{(b)} Experimental results of the
normalized intensity of the down-converted photons as a function of
the fundamental wave polarization angle.}
\end{figure}

For any light propagation direction inside a non-cubic crystal,
there are two orthogonal modes of the dielectric displacement
field $\textbf{D}_1$ and $\textbf{D}_2$, each with a different
corresponding refractive index. In uniaxial crystals, such as BBO,
these two modes are known as the \emph{ordinary} (o) and
\emph{extraordinary} (e) polarizations, while in biaxial crystals
they are known as the \emph{fast} (f) and \emph{slow} (s)
polarizations, both behaving in general as an extraordinary wave
\cite{Yariv03}.

When choosing the crystal parameters, we need to consider the
polarization of the pump beam and the down-converted photons. It is
possible to calculate the directions of the dielectric displacement
vectors $\textbf{D}_1$ and $\textbf{D}_2$ of the two linearly
polarized waves in respect to the physical axes $\{e_i\}$. However,
it is more convenient to define the photon polarizations with
respect to the \textbf{P} and \textbf{R} directions. In the BBO
crystal, the pump beam is polarized along the \textbf{R} direction,
one cone is polarized in the same direction, and the other cone is
polarized in the \textbf{P} direction. In BiBO, the pump beam should
be polarized in its fast polarization mode in order to achieve
maximal conversion efficiency, which usually differs from these
convenient directions. For the general case, we define the cartesian
coordinates of the propagation direction \textbf{T} by the unit
vector $(x,y,z)$ in the optical indicatrix system $\{e_i^0\}$. Using
the Sellmeier formula for BiBO \cite{Bohaty00}, we calculated the
wavelength dependant principal refractive indices $(n_x<n_y<n_z)$.
From them we derived the slow and fast refractive indices
\cite{Boeuf00}. Using these refractive indices, the ratios between
the components of the normal polarization modes (i.e., the
components of the unit vectors along the displacement field vectors
$\textbf{D}_{i}$) are given by \cite{Yariv03}

\begin{equation}\label{PolarizationModes}
D_{i_x}:D_{i_y}:D_{i_z}=\frac{n_x^2x}{(n_i^2-n_x^2)}:\frac{n_y^2y}{(n_i^2-n_y^2)}:\frac{n_z^2z}{(n_i^2-n_z^2)},
\end{equation}
where i stands for 'fast' or 'slow'. For our crystal parameters, the
fast polarization mode of the pump beam was calculated to be
$13.2\pm 0.1^\circ$ from \textbf{P}, as shown in Fig.
\ref{BiBOPolarizationModes}(a). We also measured this value by
rotating the pump polarization direction with a half-wave plate. At
each rotation step we took a picture of the down-converted circles.
In Fig. \ref{BiBOPolarizationModes}(b) we plot the normalized
intensity of the down-converted photons vs the polarization angle.
Setting $0^\circ$ parallel to \textbf{P}, the maximal value was
obtained at an angle of $11.6\pm 0.3^\circ$ from \textbf{P}, within
the crystal fabrication errors.

In order to calculate the polarization of the down-converted photons
at the cones intersection points, we need to consider these two
propagation direction inside the crystal. For the down-converted
photons propagating at the top left (bottom right) intersection
point in Fig. \ref{BiBOPolarizationModes}(a), the polarization
direction of the fast wave is at $14.5^\circ\,(16.0^\circ$) from
\textbf{P}. The slow polarization modes are perpendicular to the
fast modes.

\subsection{The effective second order nonlinearity}
\label{subsec:Deff}

\begin{figure}[tb]
\centering\includegraphics[width=5in]{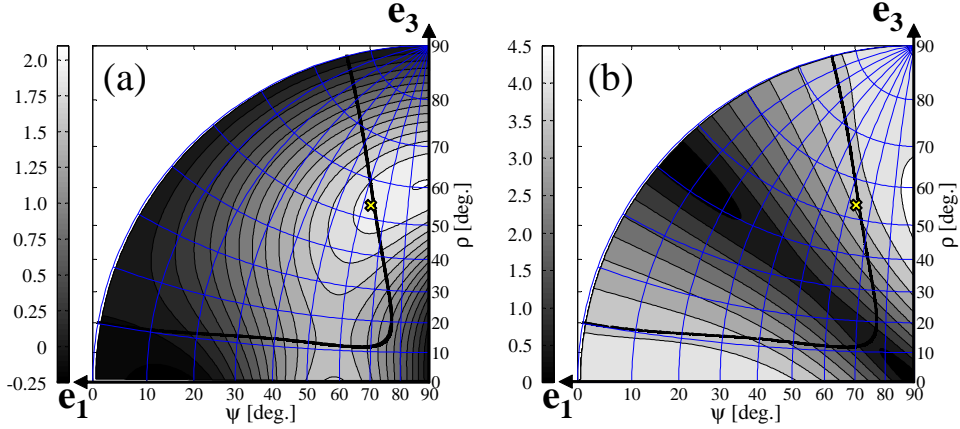}
\caption{\label{BiBO_SWO_and_d_eff} \textbf{(a)} The calculated
$d_{eff}$ [pm/V] of BiBO. Each contour line marks a step of
0.14\,pm/V. The thick black line represents collinear type-II
phase-matching directions for $\lambda_f$=390\,nm. The X symbol
marks the chosen pump direction in this work. \textbf{(b)} The
Spatial walk-off angle [deg.] for BiBO. Each contour line marks a
step of $0.45^\circ$.}
\end{figure}

The effective strength of the second order nonlinear coefficient
$d_{eff}$ is an important consideration for the crystal design.
For the uniaxial BBO crystal, there is an analytical expression
that appears in Ref. \cite{Zernike73}. Using the BBO \textbf{d}
matrix elements from Ref. \cite{Eckardt90} and our wavelength
parameters, a maximal value of $d_{eff}=1.15$\,pm/V is calculated.
The calculation assumes a collinear type-II phase-matching
process.

A rigorous treatment of biaxial crystals appears in Ref.
\cite{Petrov05}. We used the relevant formula for $d^{fsf}_{eff}$ of
collinear type-II phase-matching in the $\{e_i^0\}$ reference system
(the $fsf$ indices refer to the pump and the down-converted photon
polarization modes). The calculation results were rotated to the
$\{e_i\}$ reference system, where for the parameters used in this
work we get $d_{eff}=2.00$\,pm/V. The calculation considers a
wavelength of 780\,nm, although there is almost no wavelength
dependency. For this calculation we used the four \textbf{d} matrix
elements given in Ref. \cite{Petrov05}. The results for any
\textbf{T} direction are shown in Fig. \ref{BiBO_SWO_and_d_eff}(a).
Note that because the calculation assumes collinear propagation, the
results have significant meaning mainly in the vicinity of the
collinear phase-matching curve. Furthermore, we have also removed
the Kleinman symmetry assumption of Ref. \cite{Petrov05} and derived
a formula containing the eight \textbf{d} matrix elements given in
Ref. \cite{Ghotbi04}. This generalization resulted with a similar
value ($d_{eff}=2.02$\,pm/V). The almost doubled value of the
nonlinear parameter of BiBO compared to BBO promises a major
advantage for the generation of entangled photons.

\subsection{The spatial walk-off angle}
\label{subsec:SWO} During the propagation through a birefringent
crystal, the Poynting vector may point away from the direction
defined by the \textbf{k} vector, depending on the beam polarization
\cite{Born93}. This phenomenon is called spatial walk-off. It should
be taken into consideration when designing a polarization entangled
photon source since it can create spatial labeling of the
down-converted photons, which in turn will reduce the entanglement
quality. The spatial walk-off angle $\theta_{swo}$ between the
Poynting vector and the \textbf{k} vector, together with the crystal
thickness L, determines the overall spatial walk-off. The pump beam
spot-size at the crystal should be large compared to the spatial
walk-off in order to prevent the labeling effect \cite{Kwiat95}.

In uniaxial crystals, such as BBO, an ordinary photon's \textbf{k}
vector and Poynting vector have the same direction while an
extraordinary polarized photon deviates from that direction by an
angle that can be calculated using a simple analytical expression
\cite{Butcher90}. In biaxial crystals, such as BiBO, both the fast
and slow polarized photons deviate from the direction defined by
the \textbf{k} vector while passing through the crystal. The
spatial walk-off angle in this case is the angle between the two
down-converted photons' Poynting vectors.

We present here the results of a numerical approach for the walk-off
calculation for BiBO. The direction of the Poynting vectors of the
slow and fast down-converted photons are normal to the surface of
the corresponding indicatrix. For each photon, we calculated three
wave vectors with small deviations from their propagation direction
\textbf{k}. We then found the plane that contains these three
vectors. The direction normal to this plane is the direction of the
Poynting vector. The angle between the two Poynting vectors of the
slow and fast photons is the required walk-off angle. Note that it
is also possible to treat this problem analytically, but as our
numerical results are sufficiently accurate, we leave the rigorous
treatment for a later work.

We calculated numerically the spatial walk-off angle in BBO and BiBO
for a wavelength of $\lambda_{dc}=780$\,nm. We have validated our
numerical approach by comparing its results to the analytical
expression for BBO \cite{Butcher90}. The typical deviation between
the numerical and analytical calculations is about $10^{-6}$ degree.
For collinear PDC in BBO the walk-off angle is
$\theta_{swo}=4.15^\circ$, corresponding to an overall walk-off of
$145\,\mu$m for a 2\,mm thick crystal. The results for BiBO are
presented in Fig. \ref{BiBO_SWO_and_d_eff}(b). For our crystal
parameters, the calculated walk-off values are
$\theta_{swo}=3.55^\circ$ for one of the cones' intersection points
and $\theta_{swo}=3.6^\circ$ for the other. These results correspond
to a deviation of about $95\,\mu$m for the 1.5\,mm thick crystal
used in our experiments.

\subsection{The temporal walk-off}
\label{subsec:TWO} As their name suggests, the two polarization
modes propagate through the birefringent crystal with different
group velocities. This may cause temporal distinguishability between
the slow and fast photons. This phenomena is known as temporal
walk-off. A birefringent crystal of thickness L separates the
photons by
\begin{equation}\label{TemporalWalkOff}
\delta
T=\frac{L}{v_s}-\frac{L}{v_f}=L(\frac{n_r^s}{c}-\frac{n_r^f}{c})=L\frac{\Delta n_r}{c}\,,
\end{equation}
where $c$ is the speed of light in vacuum, and $v_s\,(v_f$) and
$n_r^s\,(n_r^f$) are the group velocity and the ray refractive
index of the slow (fast) photon, respectively. The ray refractive
index $n_r$ and the refractive index $n$ are related via
$\,n_r=n\cos\alpha$, where $\alpha$ is the angle between the
\textbf{k} vector and the corresponding Poynting
vector\cite{Born93}. The problem is more significant when $\delta
T$ is comparable to or larger than the coherence time $\tau_c$. We
addressed this issue with two methods. The first is to add two
compensating crystals, cut at the same directions as the
generating crystal but of half the thickness, in each
down-conversion path \cite{Kwiat95}. The second approach is to
overlap the two photons at a polarizing beam splitter (PBS)
\cite{Kim03}, as will be described later in Sec. \ref{subsec:ES}.

For our crystal parameters, $\Delta n_r$ is approximately 0.05 for
BBO and 0.15 for BiBO, which results with $\delta T=330$\,fs for a
2\,mm thick BBO and $\delta T=750$\,fs for a 1.5\,mm thick BiBO.
Compensation is required in both cases as these values are larger
than $\tau_c=180$\,fs, the coherence time that corresponds to the
used 3\,nm filters.

\subsection{Pump bandwidth and the entanglement quality}
\label{subsec:BP}
\begin{figure}[tb]
\centering\includegraphics[width=5in]{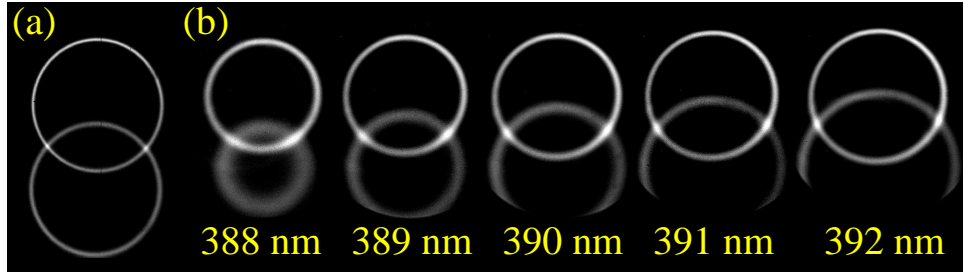}
\caption{\label{PDCPictures} \textbf{(a)} Down-converted photon
circles through a 3\,nm bandpass filter, from a 2\,mm thick BBO
crystal. \textbf{(b)} Down-converted photon circles through a 3\,nm
bandpass filter, from a 2.7\,mm thick BiBO crystal with different
pump wavelengths, as indicated. Several lower circles are cropped
due to the filter size.}
\end{figure}

One advantage of down-converting a pulsed source over a continuous
source is its energy concentration in a short coherence length
which increases the probability of higher order PDC events.
Furthermore, its timing information is inherited by the
down-converted photons. However, the pulses broadband spectrum can
cause a variety of undesired effects that decrease the
entanglement quality.

\begin{figure}[tb]
\centering\includegraphics[width=5in]{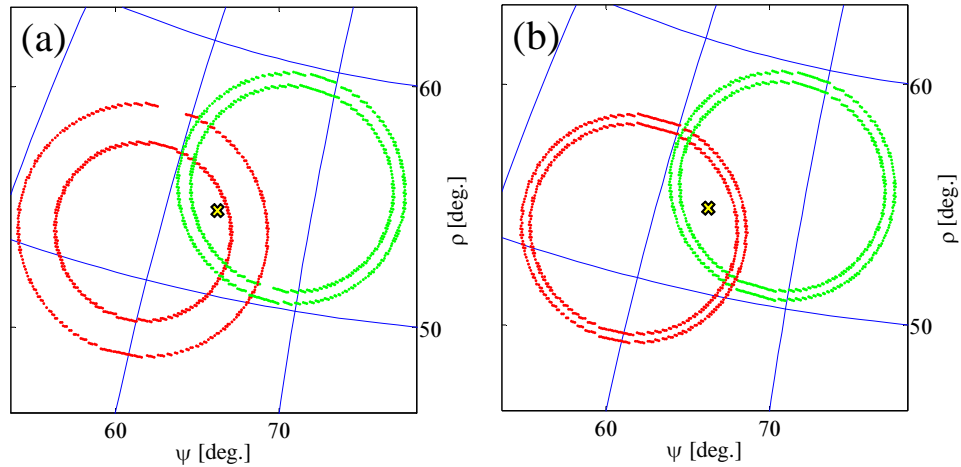}
\caption{\label{PumpAndFilter} \textbf{(a)} Stereographic projection
of non-collinear type-II PDC processes in BiBO with different pump
wavelengths. The two inner circles originate from a pump beam of
$\lambda=389$\,nm, while the two outer circles from
$\lambda=391$\,nm. The thicker ring (left, red) is polarized slow,
while the thinner one (right, green) is polarized fast. The X symbol
marks the pump direction for the BiBO crystal in this work.
\textbf{(b)} Stereographic projection of non-collinear type-II PDC
processes in BiBO with a pump wavelength of 390\,nm and different
down-converted wavelengths. The two inner circles wavelength is
781.51\,nm (left, red) and 778.5\,nm (right, green) and the two
outer circles are of the opposite process.}
\end{figure}

\begin{figure}[tb]
\centering\includegraphics[width=2.5in]{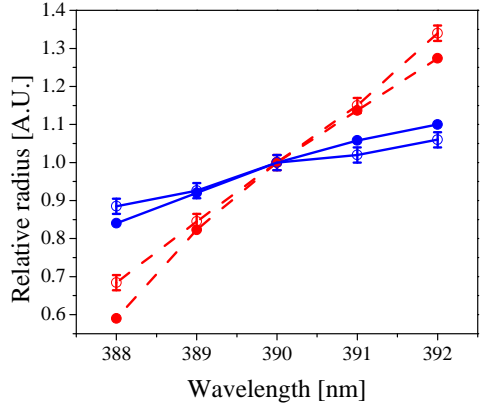}
\caption{\label{circles radii vs pump wavelength} A comparison of
the measured (open circles) and calculated (solid circles)
normalized down-converted circles radii of the slow (dashed red) and
fast (solid blue) photons.}
\end{figure}

\begin{table}[tb]
\caption{$\frac{dn}{d\lambda}$ for $\lambda=780$\,nm in nm$^{-1}$}\label{DispersionValues}
\begin{center}
\begin{tabular}{ l c c }
\hline
      & BBO                 & BiBO              \\
 slow & $3.15\times10^{-5}$ & $7.0\times10^{-5}$\\
 fast & $2.85\times10^{-5}$ & $5.0\times10^{-5}$\\
\hline
\end{tabular}
\end{center}
\end{table}

Figure \ref{PDCPictures}(a) presents a picture of the
down-converted photons from BBO recorded by a sensitive CCD camera
through a 3\,nm bandpass filter. The pump wavelength is
$\lambda_f=390$\,nm with a full width at half-maximum (FWHM) of
$\sim2$\,nm. Figure \ref{PDCPictures}(b) shows pictures of the
down-converted photons from BiBO for a few pump wavelengths. From
these pictures we note a clear difference between the widths of
the two BiBO circles, which is much smaller for BBO. For BiBO, the
lower circle, which is made out of the slow polarized photons,
changes much more than the fast polarized circle. This is due to
the difference in the dispersion of the refractive indices at a
wavelength of 780\,nm for slow and fast polarized photons in this
propagation direction. The calculated dispersion values from the
Sellmeier formulas for BBO and BiBO are presented in Table
\ref{DispersionValues}. In the BiBO case there is a $40\%$
difference, while in BBO the dispersions differs only by about
$10\%$. To ascertain these results we calculated the processes
that correspond to down-conversion of wavelengths at the FWHM
values of the pump 2\,nm spectrum
\begin{eqnarray}\label{PDCProcesses}
\nonumber fast(389\,nm)\longrightarrow  slow(780\,nm) + fast(776.01\,nm),\\
\nonumber fast(389\,nm)\longrightarrow  slow(776.01\,nm) + fast(780\,nm),\\
\nonumber fast(391\,nm)\longrightarrow  slow(780\,nm) + fast(784.01\,nm),\\
\nonumber fast(391\,nm)\longrightarrow  slow(784.01\,nm)+ fast(780\,nm).
\end{eqnarray}
We present on a stereographic projection only the circles of
$\lambda_{dc}=780$\,nm (Fig. \ref{PumpAndFilter}(a)). The circles
angular radii are measured and normalized by the radius of 780\,nm
circles from down-converting 390\,nm photons. Figure \ref{circles
radii vs pump wavelength} presents a comparison between the
numerically calculated radii and those measured from Fig.
\ref{PDCPictures}(b). The calculated (measured) slopes for the two
polarizations differ by a factor of $3.65\pm0.15$ ($2.55\pm0.05$).
The calculated slow polarization circle is thicker than the fast
polarization circle by $2.8\pm0.1$ times.

\begin{figure}[tb]
\centering{\includegraphics[width=5in]{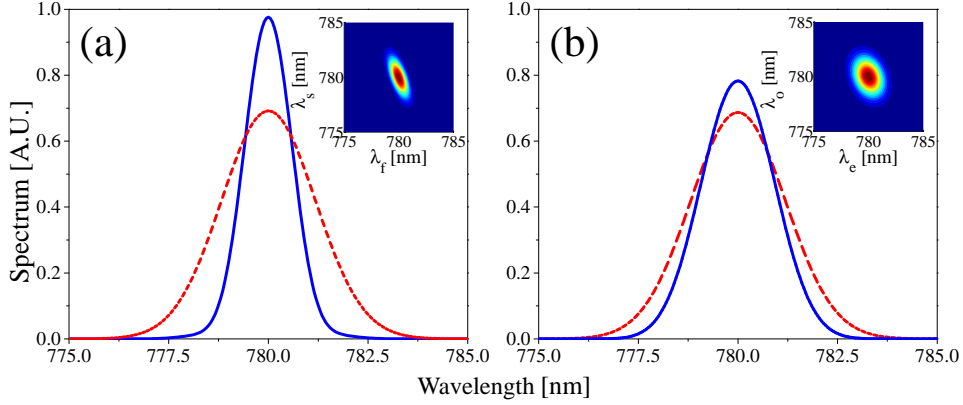}
\caption{\label{SpectraOverlap} The BiBO (\textbf{a}) and BBO
(\textbf{b}) collinear type-II PDC spectra. For BiBO (BBO), the
spectrum of the fast (extraordinary) photons is presented by a
solid blue line, while that of the slow (ordinary) photons' by a
dashed red line. In both cases, the crystals' thickness is 2\,mm,
the filter bandwidth is 3\,nm, and the pump bandwidth is 2\,nm.
Spectral overlap is $89.6\%$ for BiBO and $98.2\%$ for BBO.
\textbf{Insets:} Phase-matching spectral dependency between the
slow (ordinary) and the fast (extraordinary) photons from BiBO
(BBO). The spectra aspect ratios are 1:3 and 2:3 for BiBO and BBO,
respectively.}}
\end{figure}

\begin{figure}[tb]
\centering{\includegraphics[width=5in]{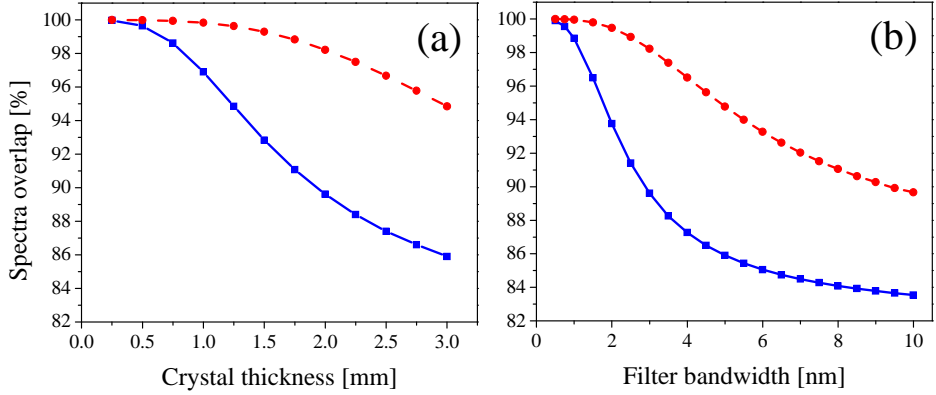}
\caption{\label{SpectraOverlapGraphs} \textbf{(a)} Spectra overlap
as a function of the crystal thickness with a 3\,nm bandpass
filter for BiBO (blue squares, solid line) and BBO (red circles,
dashed line) crystals. \textbf{(b)} Spectra overlap as a function
of the filter bandwidth for 2\,mm thick BiBO (blue squares, solid
line) and BBO (red circles, dashed line) crystals.}}
\end{figure}

In order to separate the effect of the pump bandwidth from the
effect of the filter bandwidth, we have also calculated the circle
widths due to the filter bandwidth for a 390\,nm pump. We consider
the processes that result with photons at the FWHM of the 3\,nm
filters
\begin{eqnarray}\label{PDCProcessesFilter}
\nonumber fast(390\,nm)\longrightarrow  slow(778.5\,nm) + fast(781.51\,nm),\\
\nonumber fast(390\,nm)\longrightarrow  slow(781.51\,nm) + fast(778.5\,nm).
\end{eqnarray}
The results are presented in Fig. \ref{PumpAndFilter}(b). There is
no significant effect due to the filter's width. Thus, the slow
polarized circle larger width is attributed to its higher
dispersion, that results with the asymmetry shown in Fig.
\ref{PDCPictures}(b). The filters bandwidth do not add asymmetry
between the circles. This conclusion suggests that a symmetric PDC
picture may be obtained from BiBO using a continuous pump source.

In order to evaluate the effect of these results on the quality of
the generated entangled state, we calculated the spectra of the
down-converted photons form a pulsed source in BiBO and in BBO.
Our calculation was based on the work of Grice \emph{et al.}, that
was previously applied to BBO \cite{Grice97}. As before, we used
the collinear approximation. The normalized overlap between the
two down-converted photons' spectra corresponds to the quantum
state visibility. The pump bandwidth and the crystal thickness
also influence the down-converted spectra and thus, should be
considered when designing such a polarization entangled photon
source. Bandpass filters with the proper bandwidth can enhance the
overlap between the two down-converted photons, and thus reduce
the distinguishability between them. Figure \ref{SpectraOverlap}
presents calculations of the down-converted spectra for a 2\,mm
thick BiBO and BBO crystals, assuming spectra with a FWHM of 2\,nm
for the pump photons and with 3\,nm for the bandpass filters. The
overlap between the integrated spectra of the two photons for BiBO
and BBO are $89.6\%$ and $98.2\%$, respectively. For the
entanglement measurements, we used a 1.5 mm thick BiBO crystal,
with a calculated spectral overlap of $92.8\%$. Figure
\ref{SpectraOverlapGraphs}(a) presents the dependency of the BiBO
and BBO spectra overlap on the crystal thickness for a 3\,nm
filter. The same spectra overlap as a function of the filter
bandwidth for a 2\,mm thick crystal is presented in Fig.
\ref{SpectraOverlapGraphs}(b). Although it seems as BBO can
perform better than BiBO, spectral distinguishability can be
eliminated \cite{Kim03}.

\section{Entanglement measurements}
\label{sec:ExperimentalResults}
\subsection{The experimental setup}\label{subsec:ES}
\begin{figure}[tb]
\centering\includegraphics[width=3.5in]{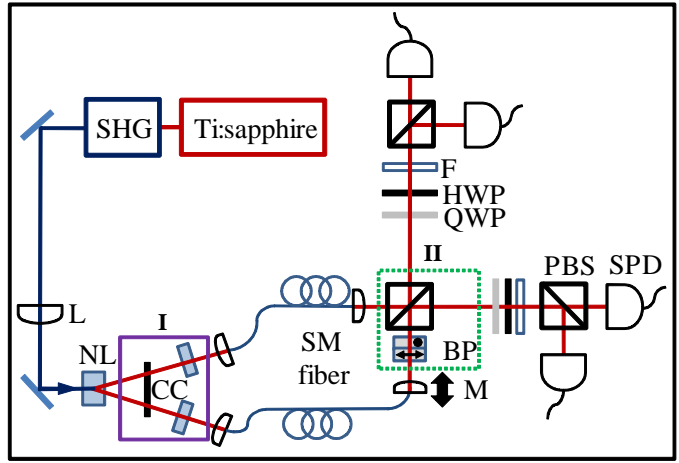}
\caption{\label{setup} The experimental setup. See text for
details.}
\end{figure}

The setup used in this experiment is presented in Fig. \ref{setup}.
The radiation of a mode-locked Ti:sapphire laser at 780\,nm is
up-converted to 390\,nm by second-harmonic generation (SHG). The
beam is focused by a lens (L) on the BBO or BiBO crystal (NL). The
spatial modes of the pump beam and the down-converted photons are
matched to optimize the collection efficiency \cite{Kurtsiefer01}.
The photons are coupled into single mode fibers (SM), where their
polarization is adjusted by polarization controllers. The relative
propagation delay between the two optical paths is adjusted by
translating one of the fiber ends with a linear motor (M). A
quarter-wave plate (QWP) and a half-wave plate at each path are used
for the quantum state tomography. The photons are spectrally
filtered by using 3\,nm wide bandpass filters (F) and coupled into
multimode fibers that guide them to the single-photon detectors
(SPD).

We tried two configurations in order to remove the temporal and
spectral distinguishability of the down-converted photons. The
elements used in each configuration are labeled I and II in Fig.
\ref{setup}. In the first configuration (I), the photon
polarizations are $90^\circ$ rotated by a half-wave plate (HWP), and
temporal and spatial walk-offs are corrected by compensating
crystals (CC) of half the thickness of the generating crystal. In
the second configuration (II), two perpendicularly oriented Calcite
crystals (arrows indicate the optical axis direction) are used for
aligning the birefringent phase (BP). The photons are then
overlapped at a PBS.

\subsection{Experimental results}
\label{sec:QEM}
\begin{figure}[tb]
\centering\includegraphics[width=5in]{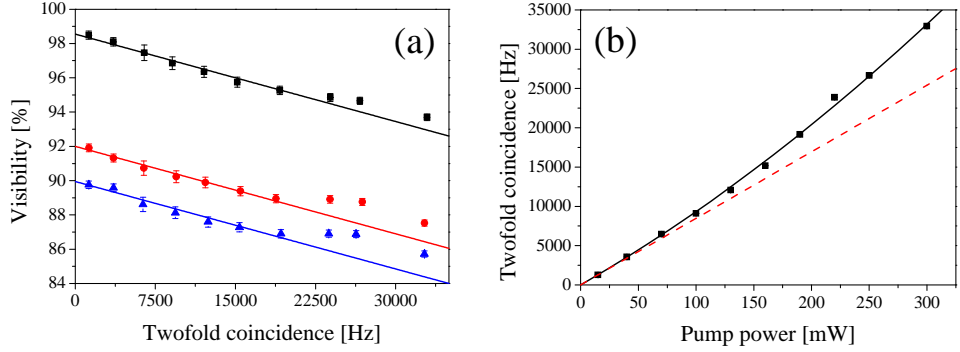}
\caption{\label{methodI} Results with configuration I. \textbf{(a)}
Visibilities vs the twofold coincidence rate in three polarization
bases: HV (black squares), PM (red circles), and RL (blue
triangles). Straight lines represent linear fits, calculated without
the last three points, where stimulation is more significant.
\textbf{(b)} Twofold coincidence rates vs pump power. The solid
black line represents the quadratic fit and the dashed red line the
linear slope at low pump powers.}
\end{figure}

\begin{figure}[tb]
\centering\includegraphics[width=5in]{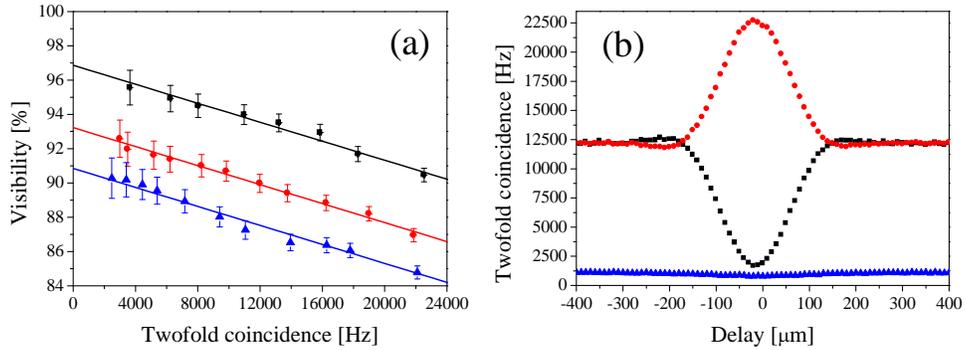}
\caption{\label{methodII} Results with configuration II.
\textbf{(a)} Visibilities vs the twofold coincidence rate in three
polarization bases: HV (black squares), PM (red circles), and RL
(blue triangles). Straight lines represent linear fits. \textbf{(b)}
Twofold coincidence rates as a function of the optical path
difference. The red circles correspond to a projection to the
$|\phi^+\rangle$ state and the black squares, a projection to the
$|\phi^-\rangle$ state. Blue triangles represent coincidence events
from the same side.}
\end{figure}

We generated polarization entangled states with a 1.5\,mm thick BiBO
crystal and compensated for distinguishability effects with two
configurations \cite{Kwiat95, Kim03} (see Sec. \ref{subsec:ES}). In
order to characterize the entanglement quality, we recorded
visibilities \cite{Kwiat95} at three polarization bases (horizontal
and vertical linear polarizations (HV), plus and minus $45^\circ$
linear polarizations (PM), and right and left circular polarizations
(RL)). Full quantum state tomography was also performed. Comparison
is made with results obtained using a 2\,mm thick BBO crystal, in a
setup optimized for its parameters.

\begin{figure}[tb]
\centering\includegraphics[width=5in]{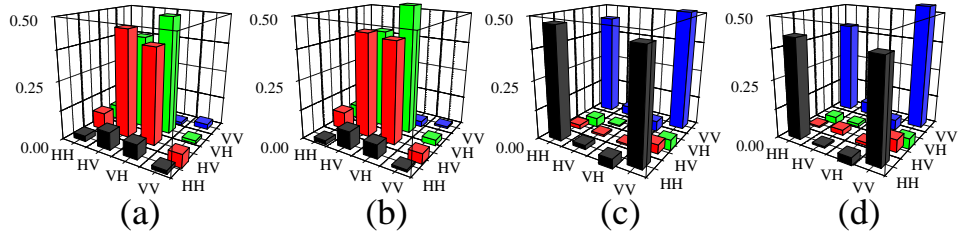}
\caption{\label{densityBiBO} Real parts of the measured density
matrices for the two configurations. Imaginary values are smaller
than 0.08 and therefore not presented. \textbf{(a)} Configuration I,
40\,mW pump. \textbf{(b)} Configuration I, 300\,mW pump.
\textbf{(c)} Configuration II, 42\,mW pump. \textbf{(d)}
Configuration II, 310\,mW pump.}
\end{figure}

Using configuration I, we generated the $|\psi^+\rangle$ Bell state.
The recorded visibilities were $V_{HV}=93.4\pm0.8\%$,
$V_{PM}=87.5\pm0.6\%$, and $V_{RL}=85.8\pm0.5\%$, with typical
twofold coincidence rates of 34000\,Hz (pump power of 300\,mW). The
detection efficiency, i.e., the ratio of the rates of coincidence
events and of single events, was $8.7\pm0.7\%$. By lowering the pump
power with a variable neutral density filter these values were
improved (see Fig. \ref{methodI}(a)). The visibilities dependence on
the pump power is due to higher order PDC events that result from
the improved generation efficiency. The extrapolated visibilities at
zero pump power are $V_{HV}=98.5\pm0.1\%, V_{PM}=92\pm0.1\%$, and
$V_{RL}=89.9\pm0.1\%$. Another manifestation of the high generation
efficiency is the stimulation process that can be seen from the
dependence of the twofold coincidence rate on the pump power (see
Fig. \ref{methodI}(b)). A quadratic function fits the data well, and
clearly deviates from the linear slope at low pump powers, a clear
signature of stimulated PDC. Density matrices were measured both at
low power (P=40\,mW, Fig. \ref{densityBiBO}(a)) and high power
(P=300\,mW, Fig. \ref{densityBiBO}(b)) \cite{James01}. Their
fidelities are $0.91\pm0.01$ and $0.88\pm0.01$, respectively,
calculated using a maximal likelihood algorithm \cite{Altepeter05}.

The second compensation scheme we have used (configuration II), was
first suggested and experimentally demonstrated by Kim \emph{et al.}
\cite{Kim03}. Using this configuration, we generated the
$|\phi^+\rangle$ Bell state. The recorded visibilities were
$V_{HV}=90.2\pm0.8\%, V_{PM}=86.9\pm0.9\%$, and
$V_{RL}=85.9\pm0.9\%$, with typical twofold coincidence rates of
22500 Hz (pump power of 310\,mW, see Fig. \ref{methodII}(a)). The
lower count rates can be attributed to the lack of spatial walk-off
compensation due to the missing compensating crystals. The detection
efficiency was $7.2\pm0.5\%$. The extrapolated visibilities at zero
pump power are $V_{HV}=96.9\pm0.1\%, V_{PM}=93.3\pm0.1\%$, and
$V_{RL}=90.9\pm0.1\%$. The density matrices for pump powers of
42\,mW and 310\,mW are shown in Figs. \ref{densityBiBO}(c) and
\ref{densityBiBO}(d), respectively. Their corresponding calculated
fidelities are $0.94\pm0.01$ and $0.90\pm0.01$.

There is another way to evaluate the entanglement quality when the
second configuration is used. Scanning the path difference before
the PBS and recording coincidences at the $45^\circ$ rotated base,
simultaneously projects on the $|\phi^+\rangle$ and $|\phi^-\rangle$
states (see Fig. \ref{methodII}(b), pump power is 320\,mW). The dip
visibility is $V_{PM}=86.4\pm0.5\%$, similar to the high pump power
visibility at the PM basis. This value is affected by contributions
from high order events. The contribution of the second order term to
the visibility can be estimated from the coincidence rate of two
orthogonally polarized photons at the same PBS output port. After
subtracting the second order events contribution, the visibility
becomes $V_{PM}=91.1\pm0.6\%$, which is comparable to the PM
visibility at low pump power.

For comparison, we have generated polarization entangled photons in
the $|\phi^{+}\rangle$ Bell state from a 2\,mm thick BBO crystal
with configuration I. The measured visibilities in the three
polarization bases $V_{HV}, V_{PM}$, and $V_{RL}$ were $95\pm1\%$,
$91\pm1\%$, and $89.5\pm1\%$, respectively. The typical twofold
coincidence rate was 37500\,Hz for a pump power of 410\,mW and
detection efficiency of $13\pm1\%$. We have also measured the
density matrix and calculated the state fidelity to be
$0.95\pm0.01$.

A meaningful comparison between BiBO and BBO should take into
account the differences between the two crystals we checked. The two
crystals were also measured in different setups, but these were
individually optimized to optimize the collection efficiency, which
depends on the crystal parameters. The BBO crystal we used was 2\,mm
thick and anti-reflection coated, while the BiBO was only 1.5\,mm
thick and uncoated. The thickness difference accounts for a factor
of 1.78, as the PDC efficiency depends quadratically on the crystal
thickness \cite{Boyd03}. The lack of coating for BiBO also accounts
for a $12\pm1\%$ loss, assuming the pump beam and the down-converted
photons hit the crystal facets perpendicularly. It should also be
considered that, due to some technical issues, we pumped the two
crystals with different powers. Thus, we calculate the
down-conversion efficiency as the number of detected pairs per
second, per mW of pump power, per mm$^2$ of crystal thickness. The
efficiency values for BiBO and BBO, as measured in configuration I,
are $58\pm1\,$Hz mW$^{-1}$mm$^{-2}$ and $23\pm1\,$Hz
mW$^{-1}$mm$^{-2}$, respectively. These values account for an
improvement by $2.5\pm0.15$, compared to the 3.09 ratio predicted by
the calculated $d_{eff}$ values of BiBO and BBO (see Sec.
\ref{subsec:Deff}).

\section{Conclusions}
We have studied the various properties of the biaxial BiBO crystal,
which are relevant for utilizing it as a polarization entangled
photon source using non-collinear type-II PDC and a pulsed pump
source. Theoretical and numerical treatment of the relevant crystal
parameters is presented. We calculated the crystal cutting angles,
the polarization directions, temporal and spatial walk-offs, and the
effective nonlinear coefficient. We have also demonstrated the
effects of crystal dispersion and the broad spectrum of the pulsed
pump on the angular and spectral properties of the down-converted
photons, and therefore on the entanglement quality. The experimental
results demonstrate the higher efficiency of BiBO compared to the
commonly used BBO, and the potential BiBO has as an ultra bright
source of entangled photons. Although it focuses on BiBO, our work
can be considered as general guidelines for considering any other
biaxial nonlinear crystal as a non-collinear type-II polarization
entangled photon source. As there are a growing number of quantum
optics experiments that require highly efficient PDC sources, we
hope that this work will encourage the use of BiBO as a source for
polarization entangled photons.

\section*{Acknowledgments}
The authors thank the Israeli Science Foundation for supporting this
work under Grants No. 366/06 and No. 546/10.

\end{document}